\documentclass[]{aa}
\usepackage{graphicx}
\usepackage{natbib}
\usepackage{amsmath}
\bibpunct{(}{)}{;}{a}{}{,}
\usepackage{txfonts}

\usepackage{umlaute}
\usepackage{amsmath}
\usepackage{multirow}

\newcommand{\dt}{${\mathrm \Delta\Theta^2}$}
\newcommand{\lt}{${\mathrm <}$}
\newcommand{\gt}{${\mathrm >}$}
\newcommand{\nums}{54}
\newcommand{\g}{${\mathrm \gamma}$}
\newcommand{\gr}{${\mathrm \gamma}$-ray}
\newcommand{\nineteen}{\mbox{1ES\ 1959+650}}
\newcommand{\twenty}{\mbox{1ES\ 2344+514}}
\newcommand{\fourteen}{\mbox{H\ 1426+428}}
\newcommand{\on}{{\mathrm{\tiny on}}}
\newcommand{\off}{{\mathrm{\tiny off}}}
\newcommand{\nS}{N$_\on$}
\newcommand{\nU}{N$_\off$}
\newcommand{\noni}{\mathrm{N_{on}^{(i)}}}
\newcommand{\noffi}{\mathrm{N_{off}^{(i)}}}
\newcommand{\sig}{$\mathrm \sigma$}

\newcommand{\fog}{$\mathrm{F_{UL}^{99\%}}$}
\newcommand{\ethr}{E$_\mathrm{thr}$}
\newcommand{\phiog}{$\mathrm{\Phi_{UL}^{99\%}}$}
\newcommand{\fluxs}{$\gamma$\,cm$^{-2}$\,s$^{-1}$}
\newcommand{\fff}{{\scriptsize$\mathrm{\Phi_{\tiny model}}$}}

\newcommand{\sm}{{\scriptsize $<$}}
\begin{document}
\title{
Observations of \nums\ Active Galactic Nuclei \\
with the HEGRA System of Cherenkov Telescopes
}
\author{ 
F.~Aharonian\inst{1}\and
A.~Akhperjanian\inst{7}\and
M.~Beilicke\inst{4}\and
K.~Bernl\"ohr\inst{1}\and
H.-G.~B\"orst\inst{5}\and
H.~Bojahr\inst{6}\and
O.~Bolz\inst{1}\and
T.~Coarasa\inst{2}\and
J.L.~Contreras\inst{3}\and
J.~Cortina\inst{2}\and
S.~Denninghoff\inst{2}\and
V.~Fonseca\inst{3}\and
M.~Girma\inst{1}\and
N.~G\"otting\inst{4}\and
G.~Heinzelmann\inst{4},
G.~Hermann\inst{1}\and
A.~Heusler\inst{1}\and
W.~Hofmann\inst{1}\and
D.~Horns\inst{1}\and
I.~Jung\inst{1}\and
R.~Kankanyan\inst{1}\and
M.~Kestel\inst{2}\and
A.~Konopelko\inst{1}\and
H.~Kornmeyer\inst{2}\and
D.~Kranich\inst{2}\and
H.~Lampeitl\inst{1}\and
M.~Lopez\inst{3}\and
E.~Lorenz\inst{2}\and
F.~Lucarelli\inst{3}\and
O.~Mang\inst{5}\and
D.~Mazin\inst{4,10}
\and
H.~Meyer\inst{6}\and
R.~Mirzoyan\inst{2}\and
A.~Moralejo\inst{3}\and
E.~Ona-Wilhelmi\inst{3}\and
M.~Panter\inst{1}\and
A.~Plyasheshnikov\inst{1\and8},
G.~P\"uhlhofer\inst{1}\and
R.~de\,los\,Reyes\inst{3}\and
W.~Rhode\inst{6}\and
J.~Ripken\inst{4}\and
G.~Rowell\inst{1}\and
V.~Sahakian\inst{7}\and
M.~Samorski\inst{5}\and
M.~Schilling\inst{5}\and
M.~Siems\inst{5}\and
D.~Sobzynska\inst{2\and9},
W.~Stamm\inst{5}\and
M.~Tluczykont\inst{4}\and
V.~Vitale\inst{2}\and
H.J.~V\"olk\inst{1}\and
C.~A.~Wiedner\inst{1}\and
W.~Wittek\inst{2}}
\offprints{M. Tluczykont, \email{Martin.Tluczykont@desy.de}}

\institute{
Max-Planck-Institut f\"ur Kernphysik,
Postfach 103980, D-69029 Heidelberg, Germany
\and
Max-Planck-Institut f\"ur Physik, F\"ohringer Ring
6, D-80805 M\"unchen, Germany
\and
Universidad Complutense, Facultad de Ciencias
F\'{\i}sicas, Ciudad Universitaria, E-28040 Madrid, Spain 
\and
Universit\"at Hamburg, Institut f\"ur
Experimentalphysik, Luruper Chaussee 149,
D-22761 Hamburg, Germany
\and
Universit\"at Kiel, Institut f\"ur Experimentelle und
Angewandte Physik,
Leibnizstra{\ss}e 15-19, D-24118 Kiel, Germany
\and
Universit\"at Wuppertal, Fachbereich Physik,
Gau{\ss}str.20, D-42097 Wuppertal, Germany
\and
Yerevan Physics Institute, Alikhanian Br. 2, 375036
Yerevan, Armenia
\and
On leave from  
Altai State University, Dimitrov Street 66, 656099 Barnaul, Russia
\and
Home institute: University Lodz, Poland
\and
Now at (2)
}

\date{Received ... / Accepted ...}

\abstract{
A sample of \nums\ selected Active Galactic Nuclei (AGN)
has been observed with the HEGRA stereoscopic system of Cherenkov Telescopes 
between 1996 and 2002 in the TeV energy regime. The observations 
were motivated by the positive results obtained for Mkn 421 and Mkn 501.
The distances of the selected objects vary over a large range 
of redshifts between $z$ = 0.004 and $z$ = 0.7. 
Among the observed AGN are the meanwhile established
TeV-emitting BL Lac type objects \fourteen\ and \nineteen.
Furthermore the BL Lac object \twenty\ and the radio galaxy M\,87 show evidence
for a signal on a 4\,$\sigma$ level. 
The observation of \twenty\ together with the Whipple results firmly 
establishes this AGN as a TeV source.
Several objects (PKS\,2155-304, BL Lacertae, 3C\,066A)
that have been claimed as TeV \gr\ emitters 
by other groups are included in this data 
sample but could not be confirmed using data analysed here.
The upper limits of several AGN included in this analysis
are compared with predictions in the frame-work of SSC models.

}
\titlerunning{HEGRA AGN observations}
\maketitle

\section{Introduction}
In the commonly adopted view the `central engine' of 
Active Galactic Nuclei (AGN) consists of a 
super massive black hole with up to 10$^9$\,${\mathrm{M_\odot}}$ surrounded 
by an accretion disk. Two relativistic plasma outflows 
(jets) perpendicular to the accretion disk 
can be observed in some AGN \citep{rees,urry}. 
So far, 
\gr s in the TeV energy regime from AGN have essentially been detected from 
objects of the BL Lac type, i.e. AGN having their jet
pointing close to the observer's line of sight. Furthermore, all known 
TeV blazars are X--ray selected BL Lac objects. 
Recently, the first detection of TeV \gr s from the radio galaxy M\,87 
with the HEGRA Cherenkov telescopes was reported \citep{m87hegra}.

{\rm Different models for the production of TeV \gr s from BL Lac
objects have been proposed. In leptonic models the 
IC mechanism is assumed to produce the TeV emission (e.g. \citet{sikora:2001}),
whereas in hadronic models the \gr s are produced 
via the interactions of relativistic protons with matter
(e.g. \citet{schlickeiser:2000}), ambient photons \citep{mannheim} or magnetic field 
\citep{aharonian:2000}, or both \citep{muecke:2001}.}

The observed TeV emission shows high 
flux variability on timescales stretching 
from months to less than an hour. 
Detailed studies
of variability of BL Lac type objects can contribute to the understanding of
their intrinsic acceleration mechanisms \citep{kraw,variability}.
The positive results obtained from the observations of the prominent
extragalactic sources of TeV \g-rays Mkn\,421 \citep{whipple1,hegra1,cat1} 
and Mkn\,501 \citep{whipple2,brad:1997:mrk,cat2} as well as their 
relevance for the question of the extragalactic background photon field 
\citep{irfelix}, 
have motivated further observations of Active Galactic Nuclei
with the HEGRA Cherenkov telescopes.
In this paper we present the results of dedicated observations of 
\mbox{\nums\ AGN} in the years 1996 to 2002
with the stereoscopic system of Cherenkov telescopes.
After a brief introduction to the HEGRA Cherenkov telescopes 
the analyzed data set will be presented followed 
by a description of the analysis
used in this paper and a presentation of the results. The paper 
closes with a discussion and a summary.

\section{The HEGRA Cherenkov telescope system}
The HEGRA stereoscopic Cherenkov telescope system (1996-2002) consisted of 
5 imaging air Cherenkov telescopes (IACTs)
\citep{daum} used in the stereoscopic observation mode
on the Canary island of 
La Palma (28.75$^\circ$\,N, 17.90$^\circ$\,W) at an altitude of 
2\,200\,m\,a.s.l. 
Additionally, one telescope (not used for 
the present analysis) was operated in stand-alone mode \citep{ct1}. 
Each system telescope was equipped with 
an 8.5 m$^2$ tesselated mirror dish of 30 single 
mirrors with a diameter of 60 cm each, and a 
camera consisting of 271 photomultiplier tubes (pixels). 
The HEGRA IACT system was operating 
at an energy threshold of 0.5~TeV for photons of vertical incidence, 
with energy and angular resolution of \mbox{$\Delta$E/E = 10 - 20 $\%$} 
and 0.1$^\circ$ respectively on an event-by-event basis. 
The field of view of each system telescope had a diameter of 4.3$^\circ$. 
The introduction of the stereoscopic observation technique results in
an improvement of the sensitivity of Cherenkov telescopes and especially
allows for an effective \g -hadron separation (see below).
The performance of the HEGRA system of Cherenkov telescopes {\rm can be found} 
in \citet{puehli}.

\section{Data Set}
Observations of \nums\ objects of the AGN class were carried out 
from 1996 to 2002, resulting in a total pre-selection exposure time 
of approx. 1150 hours (not including Mkn\,421 and Mkn\,501) 
corresponding to more than one year of continuous 
observations in moonless nights with the HEGRA IACT system. The total observation 
time accumulated for Mkn\,421 and Mkn\,501 amounts to more than 1500 hours.
The results of the HEGRA observations from these two objects were presented in 
different publications \citep{brad:1997:mrk,aha:1999:mrk2,aha:1999:mrk3,aha:1999:mrk,sam:2000:mrk,aha:2001a:mrk,aha:2001b:mrk,variability} 
and are not included in the present work.

Mkn\,421 and Mkn\,501 excluded,
the data set contains 37 objects identified and confirmed as BL Lac type objects,
10 radio galaxies, 4 Seyfert galaxies, 1 quasar and 2 galaxies following the 
catalogues of \citet{sm,pg,vc}. 
The object 1ES\,0806+524 is one of the BL Lac objects which was proposed by
\citet{tinyakov} to coincide with an AGASA UHECR triplet, thus being a candidate for
the acceleration of ultra high energy cosmic rays.
The distances of the objects vary over a large range of redshifts between 
$z$\,$\approx$\,0.004 (M\,87) and $z$\,$\approx$\,0.7 (PKS\,0219-164). 
{However, 
the expected absorption due to pair production of TeV \gr s with the
extragalactic background light (EBL) \citep{nikishov}
(\mbox{\g$_{\mathrm{TeV}}$ \g$_{\mathrm{EBL}}$ $\rightarrow 
{\mathrm e^+\mathrm e^-}$}) increases for larger redshifts, thus 
decreasing the detectability of objects located at large redshifts. 
Therefore, most of the observed objects were chosen with regard to their low
redshift (i.e.\,$z$\,$<$\,0.2).}

All observations were carried out in the so called `wobble' 
mode \citep{wobble}, tracking the telescopes with an offset of 
\mbox{0.5$^\circ$} in declination with respect to the object position,
allowing for simultaneous on- and off-source (background) observations. 
For consecutive runs with a duration of 20 min, the offset sign is reversed 
in order to avoid systematic effects due to acceptance inhomogeneities 
in the field of view. The background is estimated 
(similar to the method used in \citet{m87hegra}) using a ring segment 
concentric to the camera center
at the same radial distance to the camera
center as the on-source region (i.e. 0.5$^\circ$).
A segment with opening angle $\eta = 70^\circ$ is 
excluded from the background region in order to avoid possible 
contamination from the on-source region.
This method makes sure that on- and 
off-source measurements are both taken with identical radial camera 
acceptances and allows at the same time large
background statistics.

Two a priori cuts on the system trigger rate are applied to each run 
in order to exclude runs taken under bad weather conditions 
and to reduce systematic effects in the determination of the 
excess rate. In a first step a minimum trigger rate of \mbox{7 Hz} is required. 
This cut excludes data taken under the worst weather conditions.
In a second step, an expected rate, depending on the hardware settings and
the zenith angle of the observations 
is calculated from the parameters of a fit to all data of 
one period with constant hardware settings (all runs with trigger rate 
lower than 7~Hz are excluded from this fit).
{Runs with rates below 80\,\% of the expected rate
are rejected.}
Additionally, runs with technical problems are excluded.
After the application of the above run-selection criteria the total clean data set 
amounts to an observation time of 1017 hours.
In Table \ref{table1} all observed objects are 
listed with J2000 coordinates, redshift and object type 
ordered by ascending redshift. 
Additionally the observation time spent on each object as well as the results 
of the analysis described below are listed.

\section{Data Analysis}
The Cherenkov light generated by an air shower initiated 
by a primary \gr\ or hadronic particle is seen as 
an elliptical image in each triggered camera.
Since each telescope has a different viewing angle relative to the shower 
axis a complete geometrical reconstruction of the air shower is possible
with an image analysis of at least two telescopes.

Before the reconstruction of direction and shower core position
the following cuts are applied. Reconstructed images with more than 15 
defective camera pixels 
are rejected. 
A minimum amount of light (\emph{size}) of 40 
photo-electrons (ph.e.) is required in an image.
{Images with a distance of the center of gravity to the camera center of
more than 78\,\% of the camera radius are rejected in order to avoid truncation
by the camera border.}

After application of the above image selection criteria at least three remaining
images are required in this analysis 
for the reconstruction of the direction and 
the core impact position of an event.
This improves the quality of the reconstruction, the angular resolution 
and the separation between \gr\ and cosmic ray (hadronic background) induced 
air showers. 
The stereoscopic technique allows for an event-by-event reconstruction 
of the direction of the primary particle. 
{\rm Since the shape of the elliptical images also depends on the
shower core position}, the reconstruction
of the shower impact parameter for each telescope
provides a means of scaling the individual 
widths of the elliptical shower images of each telescope 
with expected widths for \g-ray induced shower images 
from Monte-Carlo simulations.
The mean of the scaled widths is called mscw-parameter 
and provides a very good \g -hadron separation{. 
This is described in detail in} \citet{konopelko}.
The optimum cut value for a \gr\ signal search is found to be mscw = 1.1.

For the reconstruction of the direction of the primary particle,
algorithm \# 3 from \citet{hofmann} is used.
The angle $\Delta\Theta\,\mathbf{=\,|\Theta_0 - \Theta_r|}$ 
between the object direction {$\mathbf{\Theta_0}$} and the reconstructed
shower axis {$\mathbf{\Theta_r}$} is called the angular distance.
In case of a signal from a source with point-like emission, 
the distribution of the squared angular distance \dt\ 
is expected to 
accumulate entries at { small values starting
from 0}, i.e. the signal region. The extension
and shape of the signal accumulation reflects the 
angular resolution of the system and depends on 
the telescope multiplicity, the zenith angle and the hardware setup of the 
telescope system. { Therefore, the cut on the angular distance also depends
on the parameters mentioned above.}
Using data of the well known Crab Nebula, the cut on \dt\ is { thus}
optimized individually for different hardware setups, 
multiplicities and zenith angle intervals.
This method takes the dependencies of the angular resolution described above
into account and leads to results consistent with earlier analyses.
{ Typical values of the \dt-cut are 0.008\,deg$^2$ for events reconstructed
with five triggered telescopes (having the best angular resolution)
to 0.015\,deg$^2$ for 3-telescope events.
Similarly, the cut on the core impact position slightly depends on
the zenith angle (ZA) of the observation. The optimum values found 
for this cut are 200\,m (low ZA), 400\,m (medium ZA) and 600\,m (high ZA).}
{ In Table \ref{cuttable} all selection criteria and cuts are summarized.}

Different cuts on \dt\ imply different solid angle ratios of
on- and off-source region $\alpha = \Omega_{\mathrm{on}}/\Omega_{\mathrm{off}}$ 
($\alpha$-factor) for each subset. Therefore the significance of an excess 
is calculated using a formula 
based upon the likelihood Eq.\,17 of \citet{lima} but generalized for 
data subsets with different $\alpha$-factors\footnote{In 
\citet{lima} the significance is derived from the ratio of 
the conditional probabilities for `background assumption' and 
`signal assumption'. Substituting both assumptions with a sum over 
data subsets with different $\alpha$--factors 
and a straight forward calculation leads to the above 
generalized formula.}:
\begin{equation*}
{\mathrm
S = \sqrt{2} \times
    \biggl[
  \sum_i \noni\, {\mathrm {ln}}\biggl(\frac{\sum_i \noni}{\sum_i\frac{\alpha_i}
                             {1+\alpha_i}(\noni + \noffi)}\biggr)
}
\end{equation*}
\begin{equation*}
    + \sum_i \noffi\, {\mathrm {ln}}\biggl(
\frac{\sum_i \noffi}{\sum_i\frac{1}{1+\alpha_i}(\noni+\noffi)}\biggr)\biggr]^{1/2}
\end{equation*}

The variability of each object is investigated using the Kolmogorov
and the Prahl test \citep{kolmogorov,prahl}. 
Both tests result in a significance 
for burst-like behaviour, given a time sequence of events. The Prahl test is
especially sensitive to burst-like behaviour with a small duty cycle.

For each object a Crab Nebula \g -rate as expected for identical 
observational conditions (zenith angle, hardware setup) is calculated
from data.
These expected rates are used to compute flux values and upper limits on the integral 
flux following \citet{helene}.

\begin{table}
\centering
\begin{tabular}{l|lcl}   \hline
\multirow{3}{23mm}{run selection} 	& rate           		& \gt &  7Hz 		\\
                            		& rate deviation 		& \lt &  20$\%$		\\
                            		& technical problems		& --  &			\\\hline
\multirow{3}{23mm}{image selection} 	& $\#$ of defective pixels 	& \lt &  15		\\
                              		& image size            	& \gt &  40 ph.e.	\\
                              		& distance	 		& $<$ &  0.78		\\\hline
\multirow{4}{23mm}{event selection} 	& telescope multiplicity	&$\ge$&  3 		\\
                              		& core distance         	& \lt &  f(subset)		\\
					& mscw				& \lt &	 1.1		\\
					& \dt				& \lt &  f(subset)\\\hline
\end{tabular}
\caption{Selection criteria of the analysis chain. The cuts were optimized 
individually for all data subsets using data of the well known Crab Nebula 
(see text). The distance is measured from the center of gravity of the image
to the camera center.
The entry `f(subset)' indicates that the cut depends on the data subset.}
\label{cuttable}
\end{table}

\section{Results}
\begin{figure}
  \centering
  \resizebox{\hsize}{!}{\includegraphics{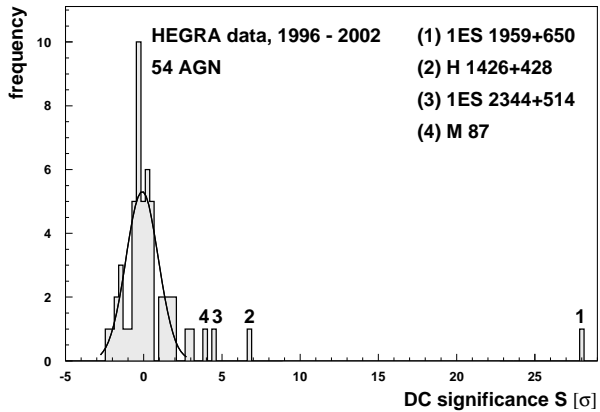}}
  \caption{Distribution of significances for all 54 objects analyzed 
           in this work. A Gaussian fit between -2.5 and +2.5 \sig\ 
           matches the core distribution very well.
         The objects \nineteen\ and \fourteen\ show a clear deviation from 
         the background expectation. 
         Further two objects, \twenty\ and M\,87 also show 
         a deviation from the background expectation 
         on the 4\,\sig\ level 
         and thus evidence for the emission of TeV \g -radiation. 
         }
  \label{distribution}
\end{figure}

A distribution of the significances for 
steady state emission (DC) of all 
analyzed objects is shown in Figure~\ref{distribution}. 
The distribution follows a Gaussian distribution of mean zero and standard
deviation one, as expected in case of a pure background sample,
with exceptions from \nineteen, \fourteen, \twenty\ and M\,87.
In Table \ref{table1}
a list of all objects analyzed in this work ordered by ascending 
redshift is shown along with their observation time and upper limits 
on the integral flux respectively flux values for the most significant objects.
The values of the calculated upper limits are found to lie between 2\,\% and
60\,\% of the Crab Nebula flux. Comparable upper limits for BL Lac objects
were found with observations of the Whipple telescope, ranging from 
6\,\% to 100\,\% of the Crab Nebula flux \citep{horan,delacalle}.
The statistical tests on burst like variability of all objects yield 
no positive results except for \nineteen \ where a high statistical 
significance ($>\,16\,\sigma$) for burst-like behaviour reflects 
the strong obvious flaring activity in May and July 2002.

The X-ray selected BL Lac object \nineteen\ \citep{einstein,schachter} was first 
reported as a TeV \gr\ emitter by the Seven Telescope Array group in 
1999 with a DC significance of 3.9 \sig\ \citep{7TA}.
HEGRA IACT system observations were carried out from July 
to September 2000, May to October 2001 and May to September 2002.
The HEGRA results on \nineteen\ were published in 
detail elsewhere \citep{1959}, the results presented in this analysis 
{ show a level of activity ranging from 0.06 to 2.9 Crab units and are} 
consistent with the earlier analysis.

\fourteen\ was reported to have a synchrotron peak
lying near or above 100 keV \citep{costamante} thus 
qualifying the object as an extreme synchrotron blazar.
Detections in the TeV energy regime from this object 
have been reported by the Whipple collaboration 
\citep{W1426} the HEGRA collaboration \citep{1426} 
and the CAT collaboration \citep{C1426}.
The observations of \fourteen\ in the years 1999, 2000 and 2002
resulted in an excess on the 7.5\,\sig\ level
\citep{aharonian:1426b},
consistent with the analysis presented here (6.6\,$\sigma$).

\begin{figure}
  \resizebox{\hsize}{!}{\includegraphics{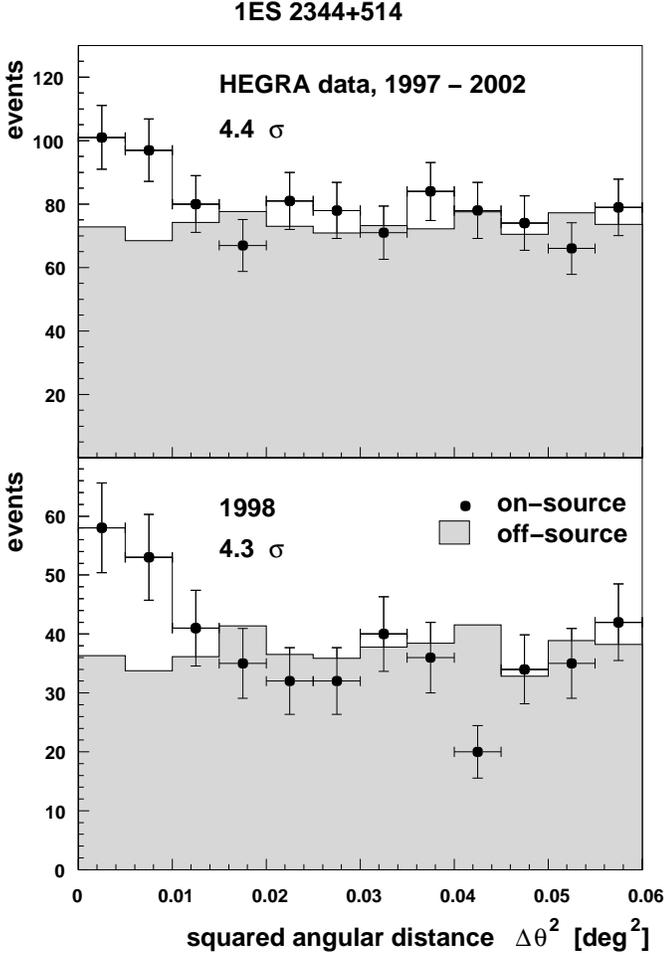}}
  \caption{Distributions of reconstructed squared angular distances { to the 
	   object direction (see text)} of the
         \twenty \ data. The distribution of the on--source events is 
         represented by the data points. The scaled 
         background (off) is shown as a shaded histogram. 
         The upper figure represents the total data set taken between 1997 and
         2002 whereas the lower figure shows the 1998 data alone.
         As can be seen from this figure the excess accumulates essentially 
         in the year 1998.
         }
  \label{2344}
\end{figure}

\twenty \ was one of the first BL Lac type objects to be reported as an
extreme synchrotron blazar with synchrotron peak energy reaching up to 100 keV
\citep{giommi}. The first TeV detection of this object 
was reported by the Whipple group in 1998 \citep{catanese}.
With an average flux of 11\,\% of the Crab Nebula flux in 1998 and 
a higher flux level of 63\,\% of the Crab flux in one night of
observations (6\,$\sigma$), the object has shown clear evidence for a 
variable flux in the Whipple data.
The results of the HEGRA observations on \twenty \ of the year 1997 and 1998 
were first reported by 
\citet{icrc99} with a DC significance of 3.3\,\sig. 
Further observations have been
carried out since the above publication.
The analysis presented here includes the complete dataset and
results in an excess of $64\,\pm\,15$ photons 
(N$_{\mathrm{on}}$ = 235, $<$N$_{\mathrm{off}}$$>$ = 171) with a significance 
of \mbox{4.4\,\sig}. The data set of
\twenty\ can be split into three independent observation periods. The first
period P1 ranges from \mbox{October to December 1997}, 
the second period P2 from 
\mbox{August to November 1998}. P1 and P2 are separated by a period of 
non observability of the object from the HEGRA site. 
Additional observations have been carried out
in September 2002 (P3).
The data subset P1 shows no evidence for a 
TeV \gr\ signal, with a DC significance
of 0.3\,\sig\ whereas the second observation period P2
yields a significance of 4.3\,\sig .
In the last observation period P3 an excess on the 2.6\,\sig \ level is found.
Tests for burst-like behaviour do not yield statistically
significant results.
In Table \ref{table2344}
the number of on-- and off--source events as well as the corresponding
significances are listed for the different data subsamples.
In Figure \ref{2344} the distributions of the reconstructed directions 
for the complete data set (P1+P2+P3) and the data set 
with the highest significance 
(P2) of \twenty\ are shown. 
\begin{table}
\centering
\caption[]{Number of on-- and off--source events and significance S for 
           the data sets from the years 1997 (P1), 1998 (P2) and 2002 (P3) on \twenty. 
           Note that the major part of the excess is accumulated in 1998.}
\begin{tabular}{clcrrc}\hline
\multicolumn{2}{c}{observation periods}		& time 	& N$_\mathrm{on}$&$\alpha$\,N$_\mathrm{off}$& S \\
		&	&$[$h$]$&\#		 &\#		  &		$[$\sig$]$ \\\hline
P1 & Oct -- Dec 1997	& 15.0  &	54	 & 52		  & 0.3	\\
P2 & Aug -- Nov 1998	& 41.8  &	128	 & 84		  & 4.3	\\
P3 & Sep 2002	        & 15.7  &	53	 & 35		  & 2.6	\\\hline
\bf $\sum$ &		& \bf 72.5  &	\bf 235	 & {\bf 171}	  & {\bf 4.4}\\\hline
\end{tabular}
\label{table2344}
\end{table}
The observed excess results in a flux of
$\Phi$(E$>$0.97\,TeV) = 
$(0.60\,\pm\,0.19)\,\cdot\,10^{-12}$\,photons\,cm$^{-2}$s$^{-1}$, 
corresponding to ($3.3\,\pm\,1.0$)\% of the Crab Nebula flux.
The errors on the flux level are dominated by the statistics
of the measurements.

In contrast to the abovementioned 4 objects the jet of the giant 
radio galaxy M87 is not aligned to our line of sight 
which makes it the only
non BL Lac type object among the 4 most 
significant objects of this data sample. 
The observed excess from the radio galaxy M87 results in a significance
of 3.9\,$\sigma$ in the present analysis and is consistent to
the results of a detailed analysis of the M87 data, 
which yield 4.1\,$\sigma$ and
were presented in a dedicated paper \citep{m87hegra}. 
After further improvements this analysis now yields 4.7\,$\sigma$ \citep{goetting}.

Several objects that were reported to be sources of TeV \gr s by other groups
are included in this data set. Among these, 
the objects 3C\,066A \citep{neshpor} and PKS\,2155-304 
\citep{chadwick,hess2155} were only observed for a very short time. 
No excess was found in the HEGRA data of these two objects.
A weak excess on the 3\,\sig\ level is found in BL-Lacertae 
which was observed 
for 29 hours, resulting in a 99\,\%\,C.L. upper limit of the order of 28\% of the Crab Nebula flux 
(see Table~\ref{table1}).
This object was reported to be a TeV \g-ray emitter by the Crimean Observatory
\citep{neshpor:bllac}.

\section{Discussion}
Predictions for \gr\ fluxes in the GeV/TeV energy regime 
have been made for several AGN by \citet{steckerdejager} and \citet{costamante2}.

In \mbox{Table \ref{tablep}} the observed integral fluxes and flux upper limits
of several objects of the present data sample are compared to predictions
made by \citet{steckerdejager}, based upon simple scaling arguments, taking
into account the Einstein Slew survey sample of BL Lacs.
The authors argue that
only high frequency peaked BL Lac objects are potential sources
of extragalactic TeV \g -radiation. 
For all objects listed in Table~\ref{tablep} the derived HEGRA upper 
limits on the integral flux were calculated for an energy threshold of
1\,TeV and for different spectral indices, assuming a power law
energy spectrum. 
The observed fluxes of \nineteen\ in its low state and 
\twenty \ are close to the predicted values. 
The calculated HEGRA upper limits for all other objects in this list
exceed the predictions. 
For some objects (e.g. 1ES\,0927+500, 1ES\,1440+122) the predicted flux levels
are a factor of 50 to 100 lower than the upper limits from HEGRA.
The level of sensitivity necessary to detect such low fluxes within a reasonable time is beyond 
the capabilities of the HEGRA Cherenkov telescopes. Such low fluxes could only be detected 
by the experiments of the next generation which have a 
higher sensitivity and a lower energy threshold.
\begin{table}
\centering
\setlength{\tabcolsep}{0.7mm}
\renewcommand{\arraystretch}{1.0}
\caption[]{Comparison of upper limits and fluxes ($\Phi$)
           derived in this work with predictions (\fff)
           made by \citet{steckerdejager}. 
           All fluxes are given in units of 
           $10^\mathrm{-12}$\,photons\,cm$^{-2}$s$^{-1}$. 
           The 99\,\%\,C.L. upper limits were
           extrapolated to a fixed energy threshold of 1\,TeV assuming
           a power law energy spectrum for 3 differential spectral indices
           $\alpha = 2.0,\,2.5,\,3.0$.}
\label{tablep}
    \begin{tabular}{lc|c|ccc}\hline

       object
     & $z$
     & {\scriptsize \fff / 10$^{-12}$}
     & \multicolumn{3}{c}{{\scriptsize $\mathrm{\Phi}/10^{-12}$}}\\

     &
     & \scriptsize \fluxs
     & \multicolumn{3}{c}{\scriptsize \fluxs}                     \\\hline

     &
     &
     & \multicolumn{3}{c}{spectral index $\alpha$}            \\

     &
     &
     & 2.0
     & 2.5
     & 3.0                      \\\hline
     I Zw 187
     &0.055
     & 0.59
     & {\scriptsize $<$}1.56
     & {\scriptsize $<$}1.51
     & {\scriptsize $<$}1.47		\\
     1ES\,2321+419
     &0.059
     & 0.14
     & {\scriptsize $<$}0.60
     & {\scriptsize $<$}0.56
     & {\scriptsize $<$}0.53		\\
     1ES\,1741+196
     &0.083
     & 0.35
     & {\scriptsize $<$}1.33
     & {\scriptsize $<$}1.29
     & {\scriptsize $<$}1.25		\\
     PKS\,2155-304
     &0.116
     & 0.88
     & {\scriptsize $<$}1.60
     & {\scriptsize $<$}3.83
     & {\scriptsize $<$}9.16		\\
     1ES\,1118+424
     &0.124
     & 0.18
     & {\scriptsize $<$}4.18
     & {\scriptsize $<$}4.12
     & {\scriptsize $<$}4.06		\\
     1ES\,0145+138
     &0.125
     & 0.26
     & {\scriptsize $<$}1.19
     & {\scriptsize $<$}1.11
     & {\scriptsize $<$}1.04		\\
     1ES\,1212+078
     &0.130
     & 0.03
     & {\scriptsize $<$}2.98
     & {\scriptsize $<$}2.86		
     & {\scriptsize $<$}2.74		\\
     1ES\,0229+200
     &0.139
     & 0.11
     & {\scriptsize $<$}2.99
     & {\scriptsize $<$}2.87
     & {\scriptsize $<$}2.75		\\
     1ES\,1255+244
     &0.140
     & 0.34
     & {\scriptsize $<$}2.03
     & {\scriptsize $<$}1.97
     & {\scriptsize $<$}1.91		\\
     1ES\,0323+022
     &0.147
     & 0.15
     & {\scriptsize $<$}0.71
     & {\scriptsize $<$}0.71
     & {\scriptsize $<$}0.71		\\
     1ES\,1440+122
     &0.162
     & 0.03
     & {\scriptsize $<$}1.37
     & {\scriptsize $<$}1.31
     & {\scriptsize $<$}1.26		\\
     1ES\,0347-121
     &0.185
     & 0.08
     & {\scriptsize $<$}7.50
     & {\scriptsize $<$}9.07
     & {\scriptsize $<$}10.96		\\
     1ES\,0927+500
     &0.186
     & 0.02
     & {\scriptsize $<$}1.02
     & {\scriptsize $<$}0.98
     & {\scriptsize $<$}0.95		\\\hline

     1ES\,2344+514
     &0.044
     & 0.80
     & 0.58
     & 0.57
     & 0.56                             \\
     1ES\,1959+650       
     &0.047
     & 2.30                       
     & \multicolumn{3}{c}{1.0 - 51.9}                \\\hline
    \end{tabular}
\end{table}

In Table \ref{comparison} the HEGRA integral flux upper limits derived 
in this work
are compared to predictions made by \citet{costamante2}.
Two different model predictions are given. The first number is taken
from a parametrization of the spectral energy distribution (SED) {originally introduced by 
\citet{fossati} and modified by \citet{donato} and \citet{costamante2} (hereafter FDC).} 
The second number is calculated using an SSC model {
from \citet{costamante2} (hereafter CG)}.
Compared to the earlier work of \citet{steckerdejager},
{ CG} introduce the new requirement of
strong radio emission for a TeV candidate source (arguing that a strong radio
emission is a good indicator for non-thermal low energy emission producing seed photons). 
Furthermore, 
several other BL Lac samples in addition to the Einstein Slew survey
sample were taken into account.
The FDC parametrization is rather suitable for predictions of high state
TeV fluxes while the SSC model predictions, designed to fit
the known synchrotron part of the SED, are more appropriate
for a quiescent state of the TeV source candidate. 
Additionally, one has to
note that along with other uncertainities
the absorption of TeV photons by the extragalactic background
radiation field was not accounted for in these models (CG).
\begin{table}
\centering
\setlength{\tabcolsep}{0.7mm}
\renewcommand{\arraystretch}{1.0}
\caption[]{Comparison of flux predictions by \citet{costamante2} ({\fff})
         with the results derived in this work. 
         All fluxes are given in units of 
         10$^\mathrm{-12}$\,photons\,cm$^{-2}$s$^{-1}$. 
         For the extrapolation of the upper limits given in Table~\ref{table1}
         a power law for 3 differential spectral indices 
         ($\alpha = 2.0,\,2.5,\,3.0$) were assumed
         and the energy threshold was fixed at \mbox{\ethr\ = 1\,TeV}.
         For 1ES\,0647+250 a redshift of 0.200 was assumed
         Two numbers for predictions above 1\,TeV are given
         by \citet{costamante2}:
         the first number is obtained from a parametrization of
         the SED adapted from \citet{fossati} and modified by \citet{donato} 
  	 and \citet{costamante} (FDC). 
         The second number results from a homogeneous one-zone 
	 SSC model described in \citet{costamante2} (CG).}
\label{comparison}
\begin{tabular}{lc|cc|ccc}\hline
        object 		& $z$ 	& \multicolumn{2}{c|}{{\scriptsize \fff$/10^{-12}$}}
							& \multicolumn{3}{c}{{\scriptsize $\Phi\,/10^{-12}$}} \\
     			& 	&\multicolumn{2}{c|}{\fluxs}& \multicolumn{3}{c}{\fluxs}\\\hline
			&	&FDC	& CG		& \multicolumn{3}{c}{spectral index $\alpha$} \\

			&	& 	& 		& 2.0  & 2.5  & 3.0    \\\hline
{\bf H 1722+119}	&0.018 & 35.2  & 0.01		& {\scriptsize $<$}3.84 & {\scriptsize $<$}3.62 {\scriptsize $<$}& 3.41   \\
{\bf 1ES 1741+196}	&0.083 & 8.4   & 0.1		& {\scriptsize $<$}1.33 & {\scriptsize $<$}1.29 {\scriptsize $<$}& 1.25   \\
	1ES 0806+524	&0.138 & 2.7   & --		& {\scriptsize $<$}4.63 & {\scriptsize $<$}4.84 {\scriptsize $<$}& 5.05   \\
	1ES 0229+200	&0.139 & 2.1   & 0.04		& {\scriptsize $<$}2.99 & {\scriptsize $<$}2.87 {\scriptsize $<$}& 2.75   \\
	RBS 0958        &0.139 & 2.8   & --		& {\scriptsize $<$}5.30 & {\scriptsize $<$}4.88 {\scriptsize $<$}& 4.50   \\
{\bf 1ES 0323+022}	&0.147 & 1.8   & --		& {\scriptsize $<$}0.71 & {\scriptsize $<$}0.71 {\scriptsize $<$}& 0.71   \\
{\bf 1ES 1440+122}	&0.162 & 2.0   & 0.1		& {\scriptsize $<$}1.37 & {\scriptsize $<$}1.31 {\scriptsize $<$}& 1.26   \\
	PG 1218+304	&0.182 & 1.5   & --		& {\scriptsize $<$}2.24 & {\scriptsize $<$}2.06 {\scriptsize $<$}& 1.88   \\
	1ES 0647+250 	&0.200 & 1.2   & --		& {\scriptsize $<$}2.61 & {\scriptsize $<$}2.31 {\scriptsize $<$}& 2.04   \\
	1ES 1011+496	&0.200 & 0.2   & --		& {\scriptsize $<$}1.84 & {\scriptsize $<$}1.85 {\scriptsize $<$}& 1.87   \\
	1ES 0120+340	&0.272 & 0.6   & --		& {\scriptsize $<$}0.72 & {\scriptsize $<$}0.66 {\scriptsize $<$}& 0.60   \\\hline
	1ES 1959+650	&0.047 & 17.4  & --		&\multicolumn{3}{c}{1.0 -- 51.9 } \\\hline
\end{tabular}
\end{table}
In addition to the flux predictions given in CG, 
we have used the FDC parametrization to calculate flux predictions for
several objects not included in the list of CG. 
{We have additionally included the effect 
of the absorption of TeV photons by pair production with 
the extragalactic background light (EBL). 
For this purpose we have used a model parametrization of 
the spectral energy distribution of the EBL
(also used in \citet{1426}, model 1)
adopted from \citet{primack} and designed to be consistent 
with our observations of known TeV Blazars. 
The luminosity distance was calculated following
\citet{ueli}.}
These predictions are compared to upper limits 
calculated in this work in Table\,\ref{others}.
\begin{table}
\centering
\setlength{\tabcolsep}{0.7mm}
\renewcommand{\arraystretch}{1.0}
\caption[]{Comparison of integral flux predictions above 1\,TeV
	   using { an own implementation of} the FDC parametrization 
           and including the absorption by the extragalactic background light 
           ({\fff}) with the 99\,\% C.L. flux upper limits ($\mathrm{\Phi}$) 
 	   derived in this work. For the objects \nineteen, 
	   \fourteen\ and \twenty\ 
	   fluxes are given. Objects printed in boldface show upper 
	   limits below the predicted values. { In case of 1ES\,0647+250
	   a redshift of 0.200 was assumed in order to be able to calculate
	   the EBL absorption.}}
\label{others}

\begin{tabular}{lc|rr|rrr}\hline
        object		& $z$ 	& \multicolumn{2}{c|}{{\scriptsize \fff $/ 10^{-12}$}}
						& \multicolumn{3}{c}{\scriptsize$\mathrm{\Phi} / 10^{-12}$} \\
     			& 	&\multicolumn{2}{c|}{\fluxs}& \multicolumn{3}{c}{\fluxs} \\\hline
			&	& FDC		& FDC    	& \multicolumn{3}{c}{spectral index $\alpha$} \\
			&	&		&+EBL		&  2.0  & 2.5 	& 3.0    \\\hline

 {\bf H\,1722+119}	& 0.018	& 36.98	&    20.53	&  \sm 3.84	& \sm 3.62	& \sm 3.41 		\\
 {\bf 1ES\,2344+514}	& 0.044	& 18.58	&     4.72	&   $=$ 0.58	& $=$ 0.57	& $=$ 0.56		\\
 {\bf Mkn\,180}		& 0.046	& 19.00	&     4.56	&  \sm 1.64	& \sm 2.00	& \sm 2.45 		\\
 1ES\,1959+650		& 0.047	& 18.01	&     4.20	&  \multicolumn{3}{c}{$= (1.0 - 51.9)$} 		\\
 {\bf 3C\,371}		& 0.050	& 12.41	&     2.80	&  \sm 2.51	& \sm 3.09	& \sm 3.81 		\\
 {\bf I Zw 187}		& 0.055	& 12.40	&     2.25	&  \sm 1.56	& \sm 1.51	& \sm 1.47 		\\
 {\bf 1ES\,2321+419} 	& 0.059	&  4.85	&     0.75	&  \sm 0.60	& \sm 0.56	& \sm 0.53 		\\
 BL-Lacertae  		& 0.069	&  1.66	&     0.22	&  \sm 4.51	& \sm 4.73	& \sm 4.96 		\\
 1ES\,1741+196		& 0.083	&  7.10	&     0.55	&  \sm 1.33	& \sm 1.29	& \sm 1.25 		\\
 PKS\,2155-304		& 0.116	&  2.40	&     0.07	&  \sm 1.60	& \sm 3.83	& \sm 9.16 		\\
 1ES\,1118+424 		& 0.124	&  2.36	&     0.05	&  \sm 4.18	& \sm 4.12	& \sm 4.06 		\\
 {\bf 1ES\,0145+13.8}	& 0.125	&  1.16	&     0.02	&  \sm 1.19	& \sm 1.11	& \sm 1.04 		\\
 {\bf 1H\,0658+595}	& 0.125	&  2.49	&     0.05	&  \sm 0.98	& \sm 1.02	& \sm 1.06 		\\
 H\,1426+428   		& 0.129	&  2.32	&     0.04	&  \multicolumn{3}{c}{$=$ 0.69} 			\\
 1ES\,1212+078 		& 0.130	&  2.74	&     0.05	&  \sm 2.98	& \sm 2.86	& \sm 2.74 		\\
 1ES\,0806+524		& 0.138	&  1.85	&     0.03	&  \sm 4.63	& \sm 4.84	& \sm 5.05 		\\
 1ES\,0229+200		& 0.139	&  2.25	&     0.03	&  \sm 2.99	& \sm 2.87	& \sm 2.75 		\\
 RBS\,0958		& 0.139	&  2.42	&     0.03	&  \sm 5.30	& \sm 4.88	& \sm 4.50 		\\
 1ES\,1255+244 		& 0.140	&  1.12	&     0.01	&  \sm 2.03	& \sm 1.97	& \sm 1.91 		\\
 {\bf 1ES\,0323+022}	& 0.147	&  1.97	&     0.02	&  \sm 0.71	& \sm 0.71	& \sm 0.71 		\\
 OQ\,530   		& 0.152	&  0.23	&  3\,$10^{-3}$	&  \sm 1.58	& \sm 1.67	& \sm 1.77 		\\
 {\bf 1ES\,1440+122}	& 0.162	&  1.69	&     0.01	&  \sm 1.37	& \sm 1.31	& \sm 1.26 		\\
 PKS\,0829+046 		& 0.180	&  0.03	&  2\,$10^{-4}$ &  \sm 0.96	& \sm 0.96	& \sm 0.96 		\\
 PG\,1218+304		& 0.182	&  1.24	&  5\,$10^{-3}$ &  \sm 2.24	& \sm 2.06	& \sm 1.88 		\\
 1ES\,0347-121 		& 0.185	&  0.83	&  3\,$10^{-3}$ &  \sm 7.50	& \sm 9.07	& \sm 10.96		\\ 
 {\bf 1ES\,0927+500} 	& 0.186	&  1.06	&  3\,$10^{-3}$ &  \sm 1.02	& \sm 0.98	& \sm 0.95 		\\
 PKS\,2254+074 		& 0.190	&  0.06	&  2\,$10^{-4}$ &  \sm 0.89	& \sm 0.85	& \sm 0.80 		\\
 MS\,0317+1834 		& 0.190	&  1.00	&  3\,$10^{-3}$ &  \sm 2.37	& \sm 2.12	& \sm 1.89 		\\
 1ES\,0647+250		& 0.200	&  0.83	&  2\,$10^{-3}$ &  \sm 2.61	& \sm 2.31	& \sm 2.04 		\\
 1ES\,1011+496		& 0.200	&  0.29	&  7\,$10^{-4}$ &  \sm 1.84	& \sm 1.85	& \sm 1.87 		\\
 1ES\,0120+340		& 0.272	&  0.37	&  8\,$10^{-5}$ &  \sm 0.72	& \sm 0.66	& \sm 0.60 		\\
 2E\,0414+0057 		& 0.287	&  0.22	&  3\,$10^{-5}$ &  \sm 2.18	& \sm 2.19	& \sm 2.20 		\\
 S5\,0716+714  		& 0.300	&  0.21	&  2\,$10^{-5}$ &  \sm 4.95	& \sm 6.22	& \sm 7.81 		\\ 
 3C\,66A    		& 0.444	&  0.11	&  1\,$10^{-7}$ &  \sm 3.29	& \sm 3.03	& \sm 2.80 		\\
 PKS\,0219-164 		& 0.698	&  0.04	&  5\,$10^{-12}$&  \sm 3.29	& \sm 4.39	& \sm 5.86 		\\ \hline
\end{tabular}                          	                                                 
\end{table}
{In the case of the FDC parametrization,
the predicted flux levels exceed the derived upper limits
for those objects printed in boldface in Table\,\ref{comparison} and Table\,\ref{others}.}
Assuming these predictions to hold true for a high state of activity,
it can be concluded that these objects were not in a flaring state
during the HEGRA observations.
{However, if we take into account the absorption by the EBL (column 4, Table\,\ref{others})
this only remains valid for four objects.}

For 1ES\,1959+650 the observed flux level during the {highest 
state of emission in the HEGRA data} is found to exceed
the predicted value by a factor of~3 {(only FDC)} to 12 {(FDC+EBL)}. 
{Given the high variability of the object 
at this time, prediction and observation can easily be accomodated.}
{The observed flux from \fourteen\ is roughly a factor of~3 below 
the value predicted by the FDC parametrisation alone. But including the absorption
by the EBL the predicted value is much lower than the observed flux (FDC+EBL).}
{The observed flux level from \twenty\ is lower than the predicted 
value by a factor of~30 {(only FDC) resp. 8 (FDC+EBL)}. 
However, this object 
has shown flux levels in the earlier Whipple data which 
exceeded the flux observed by HEGRA by a factor of 20, which shows that during HEGRA observations 
\twenty\ was indeed not in a flaring state.}
The SSC model of \citet{costamante2} predicts flux values 
well below the observed upper limits. Most of these predicted values would only be
detectable with much longer exposure times or with the next generation of Cherenkov telescopes.

\section{Summary}
A data set of 54 Active Galactic Nuclei observed with the HEGRA IACT system
between 1996 and 2002 has been analysed.
The two objects \nineteen\ and \fourteen\ are meanwhile well established
sources of TeV \g -radiation. {The evidence for TeV \gr s from \twenty\ 
is a confirmation of the detection of this object} 
by the Whipple collaboration. The detection
of TeV \gr s from M87 would be, if confirmed, the first detection of photons
in the TeV energy regime from an AGN of an object not commonly 
classified as a BL Lac object.
Upper limits have been derived for all other 50 objects.
Table~\ref{summary} summarizes the results for the most significant excesses 
seen in this data set.
{Comparisons with different model predictions indicate that a higher 
sensitivity is needed to be able to constrain the models.}
\begin{table*}
\centering
\caption[]{This Table summarizes the results of the present analysis for the most significant
           objects from the sample of 54 AGN observed with the system of stereoscopic Cherenkov telescopes.
           The number of on- and off-events as well as significance S, fluxes in units of the Crab Nebula flux (F) and
           in units of 10$^\mathrm{-12}$\,photons\,cm$^{-2}$s$^{-1}$ ($\Phi$) above the energy threshold \ethr\ of the observation are given.}
      \begin{tabular}{lrrrccc}\hline
       object           & \nS   & $\alpha$\nU & S & \ethr & F   & $\Phi$ \\
                        & $[\#]$&$[\#]$       & $[\sigma]$
                                                & $[$TeV$]$
                                                  & $[Crab]$
                                                    & $[$10$^\mathrm{-12}$\,photons\,cm$^{-2}$s$^{-1}$$]$
                                                                      				\\\hline
       1ES\,1959+650    & 1212  & 454         &28.0& 1.32& 0.06 -- 2.9 & 0.6 -- 30.5    	 \\
       H\,1426+428      & 836   & 654         &6.6 & 0.91& 0.03& 0.8    			 \\
       1ES\,2344+514    & 235   & 171         &4.4 & 0.97& 0.03& 0.6    			 \\
       M\,87            & 241   & 184         &3.9 & 0.88& 0.04& 0.8    			 \\\hline
      \end{tabular}
\label{summary}
\end{table*}

A further step towards
understanding the involved acceleration mechanisms and of the 
AGN class as a whole as well as of
the absorption by the extragalactic background radiation field
is expected from 
further observations of AGN { over a wide range of redshifts}
and especially BL Lac type objects 
with the next generation of Cherenkov telescopes { (partly already in
operation) and with future instruments}.
{ With the reduction of the energy threshold towards 100\,GeV 
the effect of the absorption by the extragalactic background light
will become less important for objects located at low redshift.
Thus, the uncertainties in the interpretation of observations
induced by this effect will be less important.}

\begin{acknowledgements}
  The support of the German Federal Ministry for Research and 
  Technology BMBF and
  of the Spanish Research Council CICYT is gratefully acknowledged. 
  GR acknownledges receipt of a von Humboldt fellowship.
  We thank the Instituto de Astrof\'{\i}sica de Canarias (IAC)
  for the use of the HEGRA site at the Observatorio del Roque de los 
  Muchachos (ORM) and for supplying excellent working conditions on La
  Palma. This research has made use of the SIMBAD database,
  operated at CDS, Strasbourg, France
  and of the NASA/IPAC
  Extragalactic Database (NED) which is operated by the Jet 
  Propulsion Laboratory, California Institute of Technology, 
  under contract with the National Aeronautics and Space
  Administration.

\end{acknowledgements}

\begin{table*}
 \centering
\caption[]{List of all objects of the HEGRA AGN data sample. The J2000 coordinates are given 
as well as redshift and object type (following \citet{sm,pg,vc}). 
The results of the analysis presented here are summarized. In case of the objects deviating 
significantly from the background expectation the flux is given in units of 
the Crab Nebula flux. For all other objects upper limits on the integral 
flux in units of the Crab Nebula flux (\fog) and in units of 
10$^\mathrm{-12}$\,photons\,cm$^{-2}$s$^{-1}$ (\phiog) are given.}
\small
\renewcommand{\arraystretch}{1.0}
\begin{tabular}{lccrr|rcrccc}\hline
object			& $\alpha_r$     &   $\delta$   & $z$ 		&type      & T 	    & \ethr   	& S &\scriptsize \fog& \scriptsize\phiog/$10^{-12}$   & F \\
                        &(hh mm ss)	 &(dd mm ss)	&               &          &$[$h$]$ &$[$TeV$]$	& $[\sigma]$ & $[Crab]$     	& \scriptsize$[$\fluxs$]$ & $[Crab]$    \\\hline
\object{1ES 0647+250}		&    06 50 46.6  &+25 03 00	&-----		&BL    	   &   4.1  & 0.78 	& 0.5 &  0.13 	&   3.35   	&		\\
\object{MG 0509+0541} 		&    05 09 26.0  &+05 41 35	&-----		&BL	   &  15.8  & 0.96 	& 1.3 &  0.11 	&   1.92   	&		\\
{\bf \object{M 87}}		&    12 30 49.4  &+12 23 28	&0.004		&F1	   &  70.0  & 0.88 	& 3.9 &        	&      	 	& 0.04         	\\    
\object{NGC 315}			&    00 57 48.9  &+30 21 08	&0.016		&F1/2      &  14.6  & 0.86 	& -0.3&  0.05 	&   1.03   	&		\\
\object{NGC 1275}		&    03 19 48.2  &+41 30 42	&0.018		&F1        &  87.6  & 0.85 	& -0.3&  0.03 	&   0.68   	&		\\
\object{H 1722+119}		&    17 25 04.5  &+11 52 15	&0.018		&BL	   &   5.1  & 0.89 	& 1.7 &  0.21 	&   4.31   	&		\\
\object{PKS 2201+04}		&    22 04 17.7  &+04 40 03	&0.028		&S1	   &  17.8  & 0.95 	& 1.7 &  0.08 	&   1.40   	&		\\
\object{V Zw 331}		&    03 13 57.0  &+41 15 37	&0.029		&BL        &   4.1  & 0.87 	& -0.2&  0.09 	&   1.93   	&		\\
\object{NGC1054}		&    02 42 15.0  &+18 13 00	&0.032		&G 	   &  57.9  & 0.86 	& -1.7&  0.02 	&   0.37   	&		\\
\object{3C 120}			&    04 33 12.0  &+05 21 15	&0.033		&F1        &  25.4  & 0.93 	& -0.7&  0.05 	&   0.86   	&		\\
\object{NGC 4151}		&    12 10 32.7  &+39 24 19	&0.033		&S1.5	   &   7.0  & 0.79 	& -0.4&  0.07 	&   1.79   	&		\\
\object{UGC01651}		&    02 09 38.5  &+35 47 51	&0.037		&G	   &  14.3  & 0.79 	& 1.3 &  0.07 	&   1.62   	&		\\
\object{UGC03927}		&    07 37 30.0  &+59 41 03	&0.041		&F2 	   &   6.3  & 1.09 	& -2.4&  0.09 	&   1.32   	&		\\
{\bf \object{1ES 2344+514}}	&    23 47 04.9  &+51 42 17	&0.044		&BL	   &  72.5  & 0.97 	& 4.4 &       	&   		& 0.03         	\\    
\object{Mkn0180}		&    11 36 26.4  &+70 09 27	&0.046		&BL	   &   9.8  & 1.50 	& -0.6&  0.12 	&   1.09   	&		\\
{\bf \object{1ES 1959+650}}	&    19 59 59.9  &+65 08 54	&0.047		&BL	   & 163.7  & 1.32 	& 28.0&       	&      		& 0.06 -- 2.9 	\\
\object{3C 371.0}		&    18 06 50.7  &+69 49 28	&0.050		&BL        &   5.4  & 1.52 	& -0.4&  0.19 	&   1.65   	&		\\
\object{4C +37.11}		&    04 05 49.3  &+38 03 32	&0.054		&S 	   &   6.7  & 0.80 	& -2.0&  0.05 	&   1.17   	&		\\
\object{I Zw 187}		&    17 28 18.6  &+50 13 10	&0.055		&BL	   &  16.0  & 0.94 	& 1.9 &  0.09 	&   1.66   	&       	\\
Cyg-A (\object{3C 405.0})	&    19 59 28.5  &+40 44 02	&0.057		&F2	   &  59.0  & 0.91 	& -0.2&  0.03 	&   0.64   	&		\\
\object{1ES 2321+419}		&    23 23 52.5  &+42 10 55	&0.059		&BL	   &  22.3  & 0.89 	& -1.6&  0.03 	&   0.67   	&		\\
\object{3C 192.0}		&    08 05 35.0  &+24 09 50	&0.060		&F2	   &   2.9  & 0.93 	& 0.3 &  0.20 	&   3.78   	&		\\
\object{4C+31.04}		&    01 19 35.0  &+32 10 50	&0.060		&FR        &   3.0  & 0.76 	& -0.3&  0.14 	&   3.83   	&		\\
{\object{BL-Lacertae}}		&    22 02 43.3  &+42 16 40	&0.069		&BL        &  26.7  & 1.10 	& 3.0 &  0.28 	&   4.10   	&		\\
\object{1ES 1741+196}		&    17 43 57.8  &+19 35 09	&0.083		&BL	   &  10.2  & 0.94 	& 0.3 &  0.07 	&   1.41   	&		\\
\object{4C+01.13}		&    05 13 52.5  &+01 57 10	&0.084		&F2 	   &   7.7  & 1.01 	& -0.2&  0.10 	&   1.73   	&		\\
\object{PKS 2155-304} 		&    21 58 52.0  &-30 13 32	&0.116		&BL	   &   1.8  & 5.72 	& 0.0 &  0.27 	&   0.28   	&		\\
\object{1ES 1118+424}		&    11 20 48.1  &+42 12 12	&0.124		&BL	   &   2.0  & 0.97 	& 0.3 &  0.24 	&   4.31   	&		\\
\object{1ES 0145+13.8}		&    01 48 29.8  &+14 02 19	&0.125		&BL        &   3.2  & 0.87 	& 1.1 &  0.06 	&   1.37   	&		\\
\object{1H\,0658+595}		&    07 10 30.1  &+59 08 20	&0.125		&BL	   &  33.7  & 1.08 	& -0.4&  0.06 	&   0.91   	&		\\
{\bf \object{H 1426+428}}	&    14 28 32.5  &+42 40 25	&0.129		&BL	   & 258.5  & 0.91 	& 6.6 &       	&          	& 0.03 		\\  
\object{3C197.1}			&    08 21 32.6  &+47 02 46	&0.130		&QSO	   &  15.0  & 0.96 	& -0.4&  0.05 	&   0.86   	&       	\\
\object{1ES 1212+078}		&    12 15 11.2  &+07 32 02	&0.130		&BL	   &   2.4  & 0.92 	& -0.6&  0.17 	&   3.24   	&		\\
\object{1ES 0806+524} 		&    08 09 49.2  &+52 18 58	&0.138		&BL	   &   1.0  & 1.09 	& -0.1&  0.29 	&   4.25   	&		\\
\object{1ES 0229+200}		&    02 32 48.7  &+20 17 17	&0.139		&BL	   &   3.0  & 0.92 	& 1.0 &  0.17 	&   3.25   	&		\\
\object{RBS 0958}		&    11 17 06.3  &+20 14 06	&0.139		&BL	   &   3.8  & 0.85 	& 2.7 &  0.28 	&   6.23   	&		\\
\object{1ES 1255+244}		&    12 57 32.0  &+24 12 39	&0.140		&BL	   &   5.9  & 0.94 	& 0.1 &  0.12 	&   2.16   	&		\\
\object{MS1019.0+5139}		&    10 22 11.0  &+51 24 00	&0.141		&S	   &  17.5  & 0.92 	& 0.1 &  0.07 	&   1.35   	&		\\
\object{1ES 0323+022}		&    03 26 13.9  &+02 25 14	&0.147		&BL	   &  14.3  & 1.00 	& -1.5&  0.04 	&   0.71   	&		\\
\object{OQ 530}			&    14 19 46.6  &+54 23 14	&0.152		&BL	   &   9.4  & 1.12 	& 0.4 &  0.10 	&   1.41   	&		\\
\object{3C 273.0}		&    12 29 06.7  &+02 03 08	&0.158		&FR        &  12.2  & 1.15 	& -0.3&  0.09 	&   1.25   	&		\\
\object{1ES 1440+122}		&    14 42 48.4  &+12 00 39	&0.162		&BL	   &  13.1  & 0.92 	& -0.9&  0.08 	&   1.49   	&		\\
\object{PKS 0829+046}		&    08 31 48.9  &+04 29 39	&0.180		&BL        &  18.0  & 1.00 	& 0.5 &  0.06 	&   0.96   	&		\\
\object{PG 1218+304}		&    12 21 22.0  &+30 10 37	&0.182		&BL	   &   3.9  & 0.84 	& -0.3&  0.12 	&   2.67   	&		\\
\object{1ES 0347-121}		&    03 49 23.0  &-11 59 26	&0.185           &BL	   &   1.9  & 1.46 	& 1.8 &  0.56 	&   5.14   	&       	       	\\
\object{1ES 0927+500}		&    09 30 37.6  &+49 50 24	&0.186		&BL	   &  13.3  & 0.94 	& 0.2 &  0.06 	&   1.08   	&		\\
\object{PKS 2254+074}		&    22 57 17.3  &+07 43 12	&0.190		&BL        &  16.3  & 0.90 	& -0.5&  0.05 	&   0.99   	&		\\
\object{MS0317.0+1834}		&    03 19 51.9  &+18 45 35	&0.190		&BL	   &   2.7  & 0.80 	& -0.5&  0.12 	&   2.96   	&		\\
\object{1ES 1011+496}		&    10 15 04.2  &+49 26 00	&0.200		&BL        &   2.0  & 1.02 	& -1.3&  0.11 	&   1.80   	&		\\
\object{1ES 0120+340}		&    01 23 08.9  &+34 20 50	&0.272		&BL	   &  18.9  & 0.83 	& -1.2&  0.04 	&   0.87   	&		\\
\object{2E 0414+0057}		&    04 16 52.5  &+01 05 23	&0.287		&BL	   &   4.5  & 1.01 	& 0.6 &  0.13 	&   2.16   	&		\\
\object{S5 0716+714} 		&    07 21 53.4  &+71 20 36	&0.300		&BL	   &   1.7  & 1.58 	& 0.7 &  0.38 	&   3.13   	&		\\
\object{3C 066A}		&    02 22 39.6  &+43 02 07	&0.444		&BL        &   1.3  & 0.85 	& -0.2&  0.17 	&   3.87   	&		\\
\object{PKS 0219-164}           &    02 22 01.0  &-16 15 16 	&0.698		&BL        &   1.7  & 1.78 	& -1.7&  0.27 	&   1.85   	&		\\\hline
\end{tabular}
 \label{table1}
\end{table*}


\small
\begin{thebibliography}{}
\bibitem[Aharonian et al.(1997)]{wobble} Aharonian, F., Daum, A., Hermann, G. 
        et al., 1997, A$\&$A, 327, L5
\bibitem[Aharonian et al.(1999a)]{aha:1999:mrk2}
        Aharonian, F., Akhperjanian, A.G., Barrio, J.A., et al., 1999a,
	A\&A 349, 11-28
\bibitem[Aharonian et al.(1999b)]{aha:1999:mrk3}
        Aharonian, F., Akhperjanian, A.G., Barrio, J.A., et al., 1999b,
	A\&A 349, 29-44
\bibitem[Aharonian et al.(1999c)]{aha:1999:mrk} 
        Aharonian, F., Akhperjanian, A.G., Andronache, M., et al., 1999c,
        A\&A 350, 757
\bibitem[Aharonian(2000)]{aharonian:2000} Aharonian, F.~A.\ 2000, New Astronomy, 5, 377
\bibitem[Aharonian et al.(2001a)]{aha:2001a:mrk}
        Aharonian, F., Akhperjanian, A.G., Barrio J.A., et al., 2001a, 
        A\&A 366, 62
\bibitem[Aharonian et al.(2001b)]{aha:2001b:mrk}
        Aharonian, F., Akhperjanian, A.G., Barrio J.A., et al., 2001b,
        A\&A 366, 746
\bibitem[Aharonian(2001)]{irfelix} Aharonian, F., 2001, Proc. of the 27th ICRC, Hamburg, Highlight Papers, 250, astro-ph/0112314
\bibitem[Aharonian(2002)]{aharonian:proton}Aharonian, F. A., 2002, MNRAS, 332, 1, 215
\bibitem[Aharonian et al.(2002a)]{variability} 
        Aharonian, F., Akhperjanian, A.G., Beilicke, M., et al., 
        2002a, A$\&$A, 393, 89
\bibitem[Aharonian et al.(2002b)]{1426} Aharonian, F., Akhperjanian, A., Barrio, J., et al., 2002b, A\&A, 384, L23 
\bibitem[Aharonian et al.(2003a)]{aharonian:1426b} Aharonian, F., Akhperjanian, A. G., Beilicke, M., et al.,
        2003a, A$\&$A, 403, 523
\bibitem[Aharonian et al.(2003b)]{m87hegra} Aharonian, F., Akhperjanian, A. G., Beilicke, M., et al., 
        2003b, A$\&$A, 403, L1  
\bibitem[Aharonian et al.(2003c)]{1959} Aharonian, F., Akhperjanian, A. G., Beilicke, M., et al., 2003c, A\&A, 406, L9 
\bibitem[Bai $\&$ Lee(2002)]{bailee} Bai, J.M. $\&$ Lee, M. G., 2002, 
        ApJ, 549, L173
\bibitem [Baltz et al.(2000)]{neutralino} Baltz, E.\,A., Briot, C., Salati, P., et al., Phys.Rev. D61 023514, 2000.
\bibitem[Bicknell $\&$ Begelmann(1996)]{m87angle} Bicknell, G. V. \& Begelmann, M. C., 1996, ApJ, 467, 597
\bibitem[Biermann et al.(2000)]{biermann} Biermann, P.\,L., Ahn, E., Medina-Tanco, G., Stanev, T., astro-ph/9911123, astro-ph/0008063 and Nucl.Phys. B Proc.Suppl. 87 (2000) 417-419
\bibitem[Bradbury et al.(1997)]{brad:1997:mrk}
        Bradbury, S. M., Deckers, T., Petry, D., et al., 1997, A\&A 320, L5
\bibitem[Catanese et al.(1998)]{catanese} Catanese, M., Akerlof, C. M., Badran, H. M., et al., 1998, ApJ, 501, 616 
\bibitem[Chadwick et al.(1999)]{chadwick} Chadwick, P. M., Lyons, K., McComb, T.J.L., et al., 1999, ApJ, 513, 161
\bibitem[Costamante et al.(2001)]{costamante} Costamante, L., Ghisellini, G., Giommi, P., et al., 2001, A\&A, 371, 512 
\bibitem[Costamante \& Ghisellini(2002)]{costamante2} Costamante, L., $\&$ Ghisellini, G., 2002, A\&A, 384, 56
\bibitem[Daum et al.(1997)]{daum} Daum, A., Hermann, G., Hess, M., et al., 1997, Astroparticle Physics, 8, 1
\bibitem[de la Calle Perez et al.(2003)]{delacalle} de la Calle Perez, I., Bond, I.H., Boyle, P.J., et al., 2003, Proc. of the 28th ICRC, Tsukuba, Vol. 5, 2571
\bibitem[Djannati-Ata{\"{\i}} et al.(1999)]{cat2} Djannati-Ata{\"{\i}}, A., Piron, F., Barrau, A., et al., 1999, A$\&$A, 350, 17 
\bibitem[Djannati-Ata{\"{\i}} et al.(2002)]{C1426} {Djannati-Ata{\"{\i}}}, A., {Khelifi}, B., {Vorobiov}, S., et al., 2002, A\&A, 391, L25 
\bibitem[Djannati-Ata{\"{\i}} et al.(2003)]{hess2155} Djannati-Ata{\"{\i}}, A., 2003, Proc. of the 28th ICRC, Tsukuba, Vol. 5, 2575
\bibitem[Donato et al.(2001)]{donato} Donato, D., Ghisellini, G., Tagliaferri, G. \& Fossati, G., 2001, A\&A, 375, 739
\bibitem[Elvis et al.(1992)]{einstein} Elvis, M., Plummer, D., Schachter, J. et al., 1992, ApJS, Vol. 80, no. 1, 257, May issue
\bibitem[Fossati et al.(1998)]{fossati} Fossati, G., Marashi, L., Celotti A., Comastri, A., Ghisellini, G., 1998, MNRAS, 299, 433
\bibitem[Ginzburg \& Syrovatskii(1965)]{ginzburg}Ginzburg, V. L. \& Syrovatskii, S. I., 1965, ARA\&A, 3, 297
\bibitem[Giommi et al.(2000)]{giommi} Giommi, P., Padovani, P., Perlman, E., et al., 2000, MNRAS, 317, 743G 
\bibitem[G\"otting et al.(2003)]{goetting}G\"otting, N. \& the HEGRA collaboration, 2003, to appear in the Proc. of the EPS 2003 conf., Aachen, also astro-ph/0310308
\bibitem[Helene(1983)]{helene} Helene, O., 1983, Nucl. Instr. Meth., 212, 319 
\bibitem[Hofmann et al.(1999)]{hofmann} Hofmann, W., Jung, I., Konopelko, A., et al., 1999, Astroparticle Physics, 12, 135
\bibitem[Horan et al.(2002)]{W1426} Horan, D., Badran, H. M., Bond, I. H., et al., 2002, ApJ, 571, 753 
\bibitem[Horan et al.(2003)]{horan} Horan, D., Catanese, M.A., Bond, I.H. et al., 2003, Proc. of the 28th ICRC, Tsukuba, Vol. 5, 2567
\bibitem[Jones et al.(1974)]{jones} Jones, T.W., O'Dell, S.L. \& Stein, W.A., 1974, ApJ, 188, 353
\bibitem[Kolmogorov(1933)]{kolmogorov} Kolmogorov, A.N., 1933, Giornale Istituto Italiano Attuari 4, 83
\bibitem[Konopelko et al.(1999a)]{konopelko} Konopelko, A., Hemberger, M., Aharonian, F., et al., 1999a, Astroparticle Physics, 10, 275 %
\bibitem[Konopelko et al.(1999b)]{icrc99} Konopelko, A., Kettler, J., and 
        the HEGRA Collaboration, 1999b, Proc. of the 26th ICRC, Salt Lake City, Vol. 3, 426 
\bibitem[Krawczynski et al.(2001)]{kraw} Krawczynski, H., Sambruna, R., Kohnle, A., et al. 2001, ApJ, 559, 187
\bibitem[Li \& Ma(1983)]{lima} Li, T. \& Ma, Y., 1983, ApJ, 272, 317 
\bibitem[Mannheim(1993)]{mannheim} Mannheim, K.\ 1993, A\&A, 269, 67  
\bibitem[Mirzoyan et al.(1994)]{ct1} Mirzoyan, R., Kankanian, R., Krennrich, F., et al., 1994, Nucl. Instr. Meth. A 351, 513 
\bibitem[M{\" u}cke \& Protheroe(2001)]{muecke:2001} M{\" u}cke, 
A.~\& Protheroe, R.~J.\ 2001, Astroparticle Physics, 15, 121
\bibitem[Neshpor et al.(1998)]{neshpor} Neshpor, Y.~I., Stepanyan, A.A., Kalekin, O.P., et al., 1998, Astr. Lett., 24, 134
\bibitem[Neshpor et al.(2001)]{neshpor:bllac} Neshpor, Y.~I., Chalenko, N.N., Stepanian, A.A et al., 2001 Astr. Rep., 45, 249
\bibitem[Nishiyama et al.(1999)]{7TA} Nishiyama, T., Chamoto, N., Chikawa, M., et al., 1999, Proc. of the 26th ICRC, Salt Lake City, Vol. 3, 370 
\bibitem[Nikishov(1962)]{nikishov} Nikishov, A.I., 1962, Sov. Phys. JETP 14, 393
\bibitem[Padovani \& Giommi(1995)]{pg} Padovani, P. \& Giommi, P., 1995, MNRAS, 277, 1477
\bibitem[Petry et al.(1996)]{hegra1} Petry, D., Bradbury, S. M., Konopelko, A., et al., 1996, A\&A 311, L13 
\bibitem[Piron et al.(2001)]{cat1} Piron, F., {Djannati-Ata{\"{\i}}}, A., Punch, M., et al., 2001, A\&A, 358, 895-906
\bibitem[Prahl(1999)]{prahl} Prahl, J., 1999, astro-ph/9909399 %
\bibitem[Protheroe et al.(2003)]{protheroe} Protheroe, R. J., Donea, A.-C., Reimer, A., 2003, Astroparticle Physics, Volume 19, 559
\bibitem[Primack(2001)]{primack} Primack, J.~R., Somerville, R.~S., Bullock, J.~S., \& Devriendt, J.~E.~G.\ 2001, High Energy Gamma-Ray Astronomy, AIP Conf. Proc., 558, 463
\bibitem[P\"uhlhofer et al.(2003)]{puehli} P\"uhlhofer, G., Bolz, O., G\"otting, N., et al. 2003, Astroparticle Physics 20, 267
\bibitem[Punch et al.(1992)]{whipple1} Punch M., Akerlof, C.W., Cawley, M.F., et al., 1992, Nature, 160, 477 
\bibitem[Quinn et al.(1996)]{whipple2} Quinn, J., Akerlof, C. W., Biller, S., et al., ApJ, 456, L83, 1996 
\bibitem[Rees(1967)]{reesSSC} Rees, M.J., 1967, MNRAS, 135, 345
\bibitem[Rees(1984)]{rees} Rees, M.J., 1984, ARA$\&$A, 22, 471 
\bibitem[Reimer(2003)]{reimer} Reimer, A., Protheroe, R.J., \& Donea, A.-C., 2003, Proc. of the 28th ICRC, Tsukuba, Vol. 5, 2631
\bibitem[Sambruna et al.(2000)]{sam:2000:mrk} Sambruna, R., Aharonian, F.A., Krawczynski, H., et al., 2000, ApJ, 538, 127
\bibitem[Schachter et al.(1993)]{schachter} Schachter, J. F., Stocke, T. J., Perlman, E., et al., 1993, ApJ, 412, 541 
\bibitem[Pohl \& Schlickeiser(2000)]{schlickeiser:2000} Pohl, M.~\& 
Schlickeiser, R.\ 2000, \aap, 354, 395
\bibitem[Sikora(2001)]{sikora:2001} Sikora, M., AIP Conf.Proc. 558 (2001) 275-288
\bibitem[Stecker et al.(1996)]{steckerdejager} Stecker, F. W., de Jager, O. C., Salamon, M. H., 1996, ApJ, 473, L75
\bibitem[Stickel et al.(1994)]{sm} Stickel, M., Meisenheimer, K. \& Kuehr, H., 1994, A\&AS, 105, 211
\bibitem[Tinyakov \& Tkachev(2001)]{tinyakov} Tinyakov, P. G. \& Tkachev, I. I., 2001, JETP Letters, 74, 445
\bibitem[Ue-Li Pen(1996)]{ueli} Ue-Li Pen, 1996, A\&A Suppl. Ser., 120, 49
\bibitem[Urry \& Padovani(1995)]{urry} Urry, C. M., Padovani, P., 1995, PASP, 107, 803 
\bibitem[V\'eron--Cetty $\&$ V\'eron(2001)]{vc} V\'eron--Cetty $\&$ V\'eron, 2001, A\&A, 374, 92 
\end{thebibliography}
\normalsize
    \end{document}